\documentclass[fleqn,twoside]{article}
\usepackage{espcrc2}
\usepackage{graphicx}
\usepackage{epsfig}

\newcommand{\BABARPubYear}    {06}
\newcommand{\BABARConfNumber} {073}
\newcommand{\SLACPubNumber}{12182}

\newcommand{\AmS}{{\protect\the\textfont2
  A\kern-.1667em\lower.5ex\hbox{M}\kern-.125emS}}
\usepackage{relsize}
\begin{document}
{\pagestyle{empty}

\begin{flushright}
SLAC-PUB-\SLACPubNumber \\
BABAR-PROC-\BABARPubYear/\BABARConfNumber \\
%\babar-PUB-\BABARPubYear/\BABARPubNumber \\
%hep-ex/\LANLNumber \\
August, 2006 \\
\end{flushright}

\par\vskip 4cm

% Title of the paper
\begin{center}
\Large \bf Review of Recent Measurements of the Sides of the CKM Unitarity Triangle
\end{center}
\bigskip

\begin{center}
\large 
Giampiero Mancinelli\\
Department of Physics,
University of Cincinnati\\  
Mail Location 11, Cincinnati, Ohio 45221-0011, USA
\end{center}
\bigskip \bigskip

% Abstract
\begin{center}
\large \bf Abstract
\end{center}
We give a review of the status of the global effort to measure the sides
of the CKM Unitarity Triangle.

\vfill
\begin{center}
Contributed to  the Proceedings of the 7$^{th}$ International Conference on
Hyperons, Charm And Beauty Hadrons\\ 2-8 July 2006,
University of Lancaster, UK
\end{center}

\vspace{1.0cm}
\begin{center}
{\em Stanford Linear Accelerator Center, Stanford University, 
Stanford, CA 94309} \\ \vspace{0.1cm}\hrule\vspace{0.1cm}
Work supported in part by
    National Science Foundation 
    Grant Number PHY-0457336 and by Department of Energy Contract
    Number DE-AC03-76SF00515.
\end{center}
\def\babar{\mbox{\slshape B\kern-0.1em{\smaller A}\kern-0.1em
    B\kern-0.1em{\smaller A\kern-0.2em R}}}
\twocolumn
\section{INTRODUCTION} 

The Cabibbo-Kobayashi-Maskawa (CKM) matrix~\cite{ckm}: 

\begin{equation}
V_{CKM} = \left( \begin{array}{ccc}
V_{ud} & V_{us} & V_{ub} \\
V_{cd} & V_{cs} & V_{cb} \\
V_{td} & V_{ts} & V_{tb} \end{array} \right)  \label{ckmmatrix2}
\end{equation}

\noindent
transforms, in the Standard Model (SM) theory, weak and mass 
eigenstates of quarks into  
each others. Its elements are proportional to
the amplitudes of such 
processes. We present the results of measurements involving
$V_{ub}$, $V_{cb}$, $V_{td}$, and $V_{ts}$, which are the smaller
ones, hence the most difficult to determine. 
As the matrix is unitary, several unitarity relations can be written,
of which the best known at the B-factories is:
${V_{ub}^\ast}{V_{ud}}+{V_{cb}^\ast}{V_{cd}}+{V_{tb}^\ast}{V_{td}}=0$. 
This can be expressed in
the imaginary plane as a triangle with sides of comparable lengths. The
apex of the Unitarity Triangle (UT) is to be constrained, in 
fact over 
constrained, to infer hints of new physics. Measurements of angles and sides
are complementary, as the former are derived from the determination of
$CP$ violating 
asymmetries (and the area of the UT is proportional to the amount of
$CP$ violation), while the latter from measuring branching fractions
(BF) of various $B$ decays, 
typically leptonic and semileptonic, or
$B$ mixing parameters. In particular, given the high precision
of the current measurement of $\sin2\beta$,  it is important to reach equal level of
precision in the measurement of the side opposite to $\beta$. All reported
results are preliminary, unless published, 
in which case appropriate journal references are given.

\section{\bf \boldmath $V_{td}/V_{ts}$}

The ratio
of the oscillation frequencies of $B_d$ and $B_s$ mesons, $\Delta m_d$ and
$\Delta m_s$, is proportional to 
$|V_{td}|/|V_{ts}|^2$. At the $B$ factories 
only $\Delta m_d$ can be measured, but the precision achieved is much
better than in the past and the world average ($\Delta 
m_d = 0.505\pm0.005$ ps$^{-1}$) is dominated by the Belle and \babar\
collaborations' measurements. 

\begin{figure}[htb]
%\vskip-1.2cm
\centerline{\psfig{file=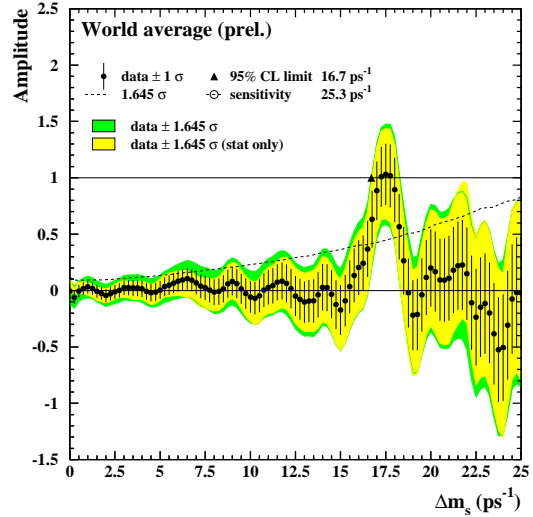,width=7.5cm}}
%\vskip-0.7cm
\caption{$B_s$ oscillation: amplitude scan versus $\Delta m_s$.}
%\vskip-0.5cm
\label{fig:fig0}
\end{figure}

Measurements of $B_s$ mixing are much more difficult as $B_s$ oscillations
are about a factor 40 
faster than those for $B_d$ mesons. $B_s$ mixing is out of the reach of the $B$
factories, but not to the energies 
of Fermilab. As the mixing is so fast, a clever way to measure $\Delta m_s$
was devised many years ago: it involves scanning the amplitude versus
proper time at 
different $\Delta m_s$ values. The true $\Delta m_s$ would give an
amplitude of about 1. Such a value 
was in fact reached years ago for $\Delta m_s\sim$17 ps$^{-1}$, but
unfortunately these 
measurements were only sensitive up to $\Delta m_s\sim$14 ps$^{-1}$, thus only
allowing a lower limit 
on $\Delta m_s$. A peak at 1 is clearly visible in the
recent CDF and D0 combined measurements (see
Fig.~\ref{fig:fig0}~\cite{hfag}) and a value
of $\Delta m_s=17.33^{+0.42}_{-0.21}(stat.)\pm0.07(syst.)$ is
achieved~\cite{deltams}.  
The precision of the measurement of $\Delta m_s$ is about 3\%, hence the
uncertainty on $|V_{ts}|/|V_{td}|=0.208^{+0.008}_{-0.007}$ is only
$\sim$4\%, as many QCD 
corrections cancel in the ratio. Fig.~\ref{fig:fig1} shows
the impact of this measurement on the UT fit~\cite{ckmfitter}.

\begin{figure}[htb]
%\vskip-0.5cm
\centerline{\psfig{file=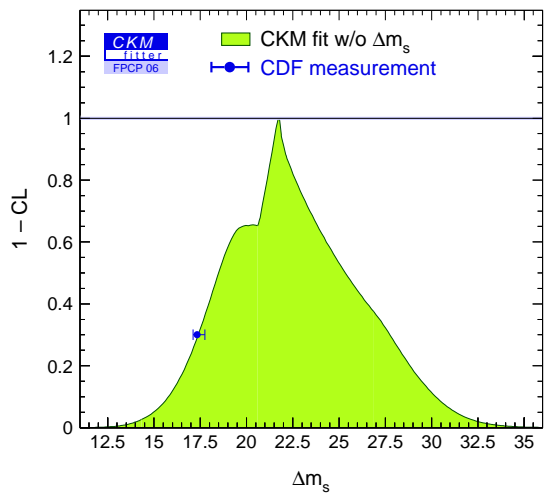,width=7.5cm}}
%\vskip-1.0cm
\caption{CDF $\Delta m_s$ measurement result versus the CKM fit without such result.}
%\vskip-0.4cm
\label{fig:fig1}
\end{figure}

A complementary measurement of $|V_{td}|/|V_{ts}|$ can be achieved using $b\to
d\gamma$ and 
$b\to s\gamma$ processes, for which  $B\to \rho (\omega) \gamma$ and $B\to
K^*\gamma$ decays can be used. 
The Belle collaboration has detected a $b\to d\gamma$ signal, hence measured
$|V_{td}|/|V_{ts}| =
0.199^{+0.026}_{-0.025}(exp.)^{+0.018}_{-0.015}(theo.)$~\cite{belle1},  
while the \babar\ collaboration has 
set a limit
$|V_{td}|/|V_{ts}|<0.19$ at 90\% confidence level
(C.L.)~\cite{babar1}. The two results are in agreement with each other and
with the $\Delta m_s$ 
results. The fit using all modes gives $0.16\pm0.02$, about two standard
deviations ($\sigma$) away from the $\Delta m_s$ result.  

\section{\bf \boldmath $V_{ub}$ AND $V_{cb}$}

\subsection{Introduction}

$V_{ub}$ and $V_{cb}$ are typically derived from
semileptonic $B$ decays. The various techniques used can be categorized
as either inclusive (when the hadron is not reconstructed) or exclusive (when
the hadron is fully reconstructed). The inclusive methods are more 
efficient but have poor signal to background ratio. The exclusive ones
have lower efficiency due to the full reconstruction of the
event, but good signal to noise ratio. Help can come from the
reconstruction of the other side (tagging side) of the event. In both cases it
is hard to discern $b\to u$ from $b\to c$ transitions 
because of the very different rates, as $V_{cb}$ is much larger than $V_{ub}$. 
Furthermore, all methods incur theoretical difficulties
when attempting to extract parton level quantities from hadron
level ones.

\subsection{Inclusive measurements}

The lepton  
momentum spectrum is harder for $b\to u l \nu$ than for $b\to c l \nu$
decays. At first order, the BF of $b \to u l \nu$ is proportional to
$|V_{ub}|^2$. 
According to the Operator Product Expansion (OPE), QCD corrections need to
be considered. These corrections are both 
perturbative (known to $\alpha^2$) and non-perturbative (although
suppressed by $1/m_b^2$). The dominant uncertainty comes from the $b$
mass, known to 
1\%. Furthermore, as $V_{ub}$ is small, the signal is practically invisible,
hence  the total rate cannot be measured. Measurements of partial rates
can instead be attempted, with the help of ad hoc kinematical selections
on such 
quantities as the lepton energy ($E_l$), the transferred momentum ($q$)
and the hadron 
system mass ($m_x$). These rates are more dependent on
non-perturbative effects and on the knowledge of the Shape Functions (SF). 

$V_{cb}$ can be determined from $b\to c l \nu$ decays as the partial
semileptonic 
rates are related to $V_{cb}$ and can be determined with the help of
theory. The lepton energy and hadron mass spectra have been measured,
together with their moments. Fig.~\ref{fig:fig2} shows the hadron
mass spectra as measured by the
Belle collaboration~\cite{vcb1}.

\begin{figure}[htb]
%\vskip-0.3cm
\centerline{\psfig{file=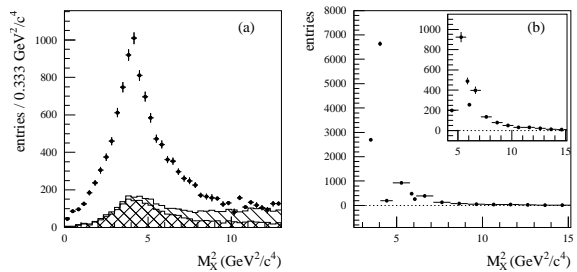,width=7.5cm}}
%\vskip-0.5cm
\caption{Hadron mass spectra (measured, (a), and unfolded (b)) in $b\to c
  l \nu$ decays from the Belle collaboration.} 
%\vskip-0.5cm
\label{fig:fig2}
\end{figure}

OPE predicts the partial rates and the moments as a function of $V_{cb}$,
the $b$ and $c$ quark masses, and other non-perturbative parameters. As each
observable has a different dependence on these quantities, a global fit
can be attempted~\cite{bf}. Furthermore $b\to s \gamma$ can
help as 
the energy spectrum of the photon is directly connected to the SF.  
The global fit gives
$|V_{cb}|=(41.96\pm0.23(exp.)\pm0.35(theo.)\pm0.59(semilept.BR.))10^{-3}$.
Other results are shown in Table~\ref{table:tab1}; note
that the error on $|V_{cb}|$ is only 
of the order of 2\%. Many measurements are used in the fit, of which the
most recent ones are from the \babar\ and DELPHI collaborations~\cite{vcb2}.
$m_b$ and  
$m_\pi^2$ are also useful in the determination 
of $V_{ub}$. Measurements from $b\to c l \nu$ and $b\to s \gamma$ processes are in
good agreement. 

\begin{table}[htb]
%\vskip-0.3cm
\caption{Results of the global OPE fit (see text). The first error is
  experimental, the second theoretical.}
\label{table:tab1}
\newcommand{\m}{\hphantom{$-$}}
\newcommand{\cc}[1]{\multicolumn{1}{c}{#1}}
\begin{tabular}{@{}ll}
\hline
Quantity           & Fit Result  \\
\hline
$m_b$                   & $4.590\pm0.025\pm0.030$ GeV  \\
$m_c$                   & $1.142\pm0.037\pm0.045$ GeV  \\
$m_\pi^2$               & $0.401\pm0.019\pm0.035$ GeV$^2$  \\
$BR_{cl\nu}$            & $10.71\pm0.10\pm0.08$\%  \\
\hline
\end{tabular}
%\vskip-0.5cm
\end{table}

A way to isolate $b\to u l \nu$ decays is to use the lepton endpoint,
i.e. selecting high energy leptons. A very accurate $b \to c l \nu$ background
subtraction is needed. Several measurements exploit this
technique~\cite{end}. The \babar\ analysis is performed in the  region
between 2.0 and 2.6
GeV (to avoid $b\to c l \nu$ background on the low side and $e^+e^-\to
u\bar u,d\bar d,s\bar s,c\bar c$ decays on the high side).
$V_{ub}$ can be extracted using the BLNP calculations~\cite{blnp}. The
results are shown in Table~\ref{table:tab2}; the
theoretical errors include the ones 
from the SF knowledge (which uses the OPE fit results).

\begin{table}[htb]
%\vskip-0.5cm
\caption{Results from lepton endpoint analyses. SF
  uncertainties are taken from the results of the OPE global fit. The
  first error is 
  experimental, the second theoretical.}
\label{table:tab2}
\newcommand{\m}{\hphantom{$-$}}
\newcommand{\cc}[1]{\multicolumn{1}{c}{#1}}
\begin{tabular}{@{}lll}
\hline
           & $E_l$ range ($GeV$) & $|V_{ub}| (10^{-3})$ \\
\hline
\babar\                 & 2.0-2.6 & $4.41\pm0.29\pm0.31$  \\
Belle                   & 1.9-2.6 & $4.82\pm0.45\pm0.30$  \\
CLEO                    & 2.2-2.6 & $4.09\pm0.48\pm0.36$  \\
\hline
\end{tabular}
%\vskip-0.5cm
\end{table}

One of the $B$ mesons can also be tagged in its fully reconstructed
hadronic decays, while identifying semileptonic decays in the recoil
system. Using selection criteria on the kinematic quantities previously
described, $V_{ub}$ can be extracted~\cite{vub1}.     

A list of all inclusive $|V_{ub}|$ measurements~\cite{hfag} is shown in
Fig.~\ref{fig:fig3}. The average, using BNLP,
is $(4.45\pm0.20(exp.)\pm0.26(theo.))10^{-3}$~\cite{vubi}.  
The experimental error is now smaller
than the theoretical. SF and weak
annihilation uncertainties should diminish with more data as well. A new
approach by Andersen and Gardi~\cite{ag} gives an even
smaller theoretical error.  

\begin{figure}[htb]
\centerline{\psfig{file=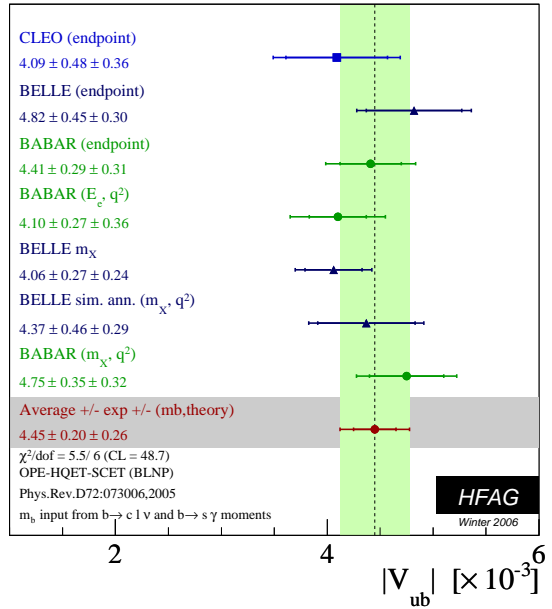,width=8.0cm}}
%\vskip-0.6cm
\caption{Inclusive measurements of $|V_{ub}|$.}
%\vskip-0.5cm
\label{fig:fig3}
\end{figure}

A novel approach realized at \babar\ uses a method by Leibovich, Low, and
Rothstein~\cite{llr}
to combine $b\to u l \nu$ and $b\to s \gamma$ measurements in order for the SF
contributions to cancel out, hence reducing the model dependence of this
measurement. A technique using weight functions had been previously proposed
by Neubert~\cite{neub}. 
Hadronically tagged and background subtracted events are used to combine the
integrated hadronic mass spectrum below a certain cut with the
high energy 
end of the measured $b\to s \gamma$ photon energy spectrum. This is
equivalent to 
trading some of the statistical error for non-perturbative theoretical
uncertainty. The 
optimal choice for the $m_x$ cut is at 1.67,
for which it is possible to obtain  72\% acceptance and
$|V_{ub}|=(4.43\pm0.38(stat.)\pm0.25(syst.)\pm0.29(theo.))10^{-3}$, which
can be 
compared  with the (worse) OPE result obtained from almost the full
spectrum:
$|V_{ub}|=(3.84\pm0.70(stat.)\pm0.30(syst.)\pm0.10(theo.))10^{-3}$~\cite{modelind}.

\subsection{Exclusive measurements}

For these measurements the Form Factors (FF) are needed, and though they are
theoretically calculable at kinematical limits, empirical extrapolations
are still necessary to extract $V_{ub}$ and $V_{cb}$. For $B\to \pi l \nu$
the FF can be calculated using several theoretical 
models~\cite{pilnuff}.  
For $B \to D^* l \nu$ the FF can be expressed as a function of $s$, the
$D^*$ boost in the $B$ rest frame. The expression depends on the
theoretical 
parameters $\rho^2$, $R_1$ and $R_2$ and the (experimental variables)
angle between the $D$ (lepton) in the $D^*$ (virtual $W$) rest frame and
the direction of the $D^*$ (virtual $W$) in the $B$ rest frame, and the
dihedral angle between the plane formed by the 
$D-D^*$ and the one formed by the $W-l$ systems~\cite{dslnuff}. The
results of the fit of the 
experimental distributions are:
$R_1=1.396\pm0.060(stat.)\pm0.044(syst.+theo.)$,
$R_2=0.885\pm0.040(stat.)\pm0.026(syst.+theo.)$,
$\rho^2=1.145\pm0.059(stat.)\pm0.046(syst.+theo.)$,
$|V_{cb}|=(37.6\pm0.3(stat.)\pm1.3(syst.)^{+1.5}_{-1.3}(theo.))10^{-3}$.  
\babar's analysis has improved the knowledge of
$R_1$ and $R_2$ by a 
factor of 5 with respect to previous CLEO measurements~\cite{r1r2}. 

Fig.~\ref{fig:fig4} reports all the exclusive $V_{cb}$ measurements to
date. The average value of
$|V_{cb}|=(40.9\pm0.9(exp.)\pm1.5(theo.))10^{-3}$ is in agreement
with  
inclusive results, although the $\chi^2/dof$ of all exclusive measurements is
quite poor (30.2/14).

\begin{figure}[htb]
%\vskip-0.5cm
\centerline{\psfig{file=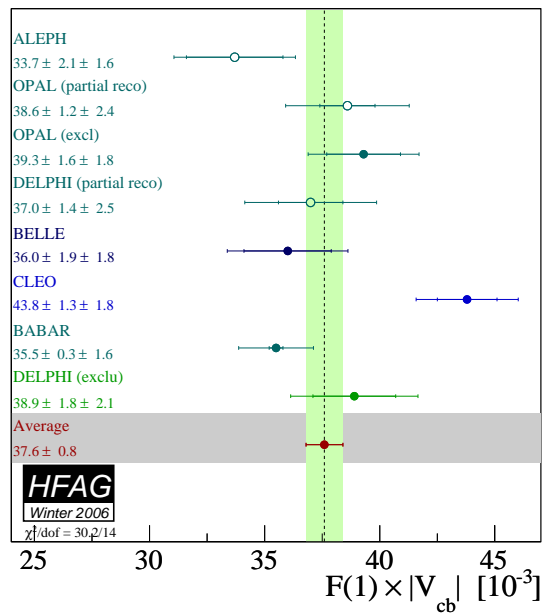,width=8.0cm}}
%\vskip-0.6cm
\caption{Exclusive measurements of $|V_{cb}|$.}
%\vskip-0.5cm
\label{fig:fig4}
\end{figure}

$V_{ub}$ can also be determined via exclusive measurements, though these
suffer from low statistics. Reconstructing $B\to \pi l \nu$ without tagging the
other side is a technique, where the neutrino information is inferred from the
event total missing 4-momentum~\cite{pilnunotag}.
Two quantities are used to discriminate between signal and background: the
beam-energy-substituted mass  
$m_{ES} \equiv \sqrt{(E_i^{*2}/2 +
  \mathbf{p}_i\cdot\mathbf{p}_B)^2/E_i^2-p_B^2}$ 
and the energy difference $\Delta E\equiv E^*_B-E_i^*/2$, 
where the subscripts $i$ and $B$ refer to the initial $e^+e^-$ system and the 
$B$ candidate respectively, and the asterisk denotes the center of mass frame. 
Using these variables, a
fit is performed in bins of $q^2$. By extracting
the $q^2$ spectrum and 
comparing it with theoretical predictions, it is possible to discriminate
between some FF models, as shown in Fig.~\ref{fig:fig5} for \babar\ data,
where, for example, the LQCD/LCSR model is clearly favored over the
ISGWII~\cite{isgw2}. 

\begin{figure}[htb]
%\vskip-0.3cm
\centerline{\psfig{file=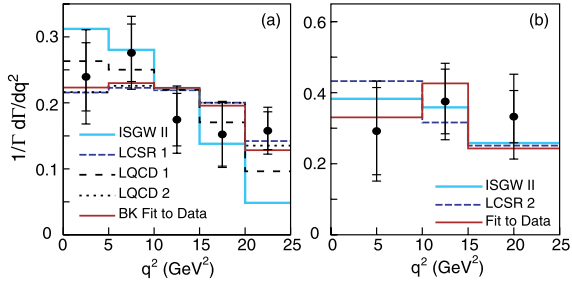,width=7.5cm}}
%\vskip-0.5cm
\caption{Measured $q^2$ spectrum ((a) $B\to \pi l \nu$, (b) $ B\to \rho l \nu$)
compared with theoretical predictions, from the \babar\ collaboration.}
%\vskip-0.5cm
\label{fig:fig5}
\end{figure}

Another procedure consists in reconstructing a semileptonic decay on the
tagging side and identifying
a $\pi l \nu$ candidate on the signal side~\cite{sltag}. Though the
$D^* l \nu$ BF is 
large, the presence of two neutrinos complicates these
measurements. By requiring the two B mesons to be back to back, it has been
possible to obtain the following BF measurements for $B^0\to \pi^- l^+
\nu$~\cite{charge}: $(1.38\pm0.19(stat.)\pm0.14(syst.))10^{-4}$ (Belle),
$(1.03\pm0.25(stat.)\pm0.13(syst.))10^{-4}$ 
(\babar); and for 
$B^+\to \pi^0 l^+ \nu$: $(0.77\pm0.14(stat.)\pm0.08(syst.))10^{-4}$ (Belle),
$(1.80\pm0.37(stat.)\pm0.23(syst.))10^{-4}$ (\babar). 
Techniques using hadronic tags yield lower statistics, but require only one
neutrino reconstruction. Furthermore they allow for high phase-space
acceptance, 
hence reduced model dependence. The \babar\ collaboration  reports $BF(B \to
\pi^0 l \nu) =
(1.28\pm0.23(stat.)\pm0.16(syst.))10^{-4}$~\cite{hadtag}. Tagged  
exclusive measurements are now becoming competitive with untagged ones.

In summary, Fig.~\ref{fig:fig8} reports all the exclusives measurements of
$|V_{ub}|$, 
given the input of several FF models and the world average exclusive BF for
$B\to \pi l \nu$ of $(1.34\pm0.08(stat.)\pm0.08(syst.))10^{-4}$. Note that
the experimental uncertainties are now 
competitive with those for inclusive measurements. The precision of the
exclusive measurements is limited mostly by the FF uncertainties
and the exclusive and inclusive measurements agree within such precision. 

\begin{figure}[htb]
%\vskip-0.5cm
\centerline{\psfig{file=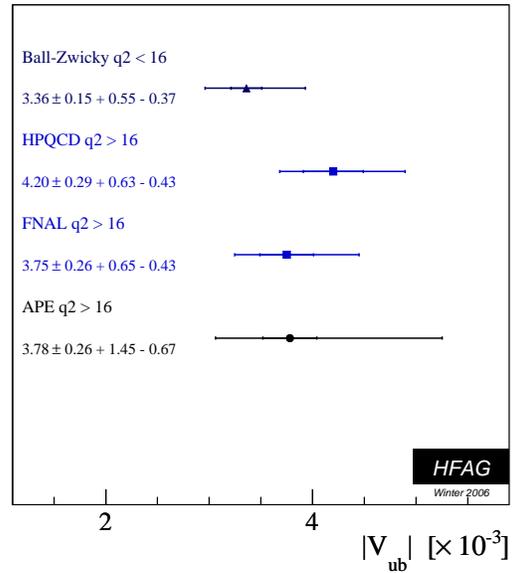,width=7.5cm}}
%\vskip-0.7cm
\caption{Exclusive measurements of
$|V_{ub}|$ for several theoretical models.}
%\vskip-0.6cm
\label{fig:fig8}
\end{figure}

\subsection{\bf \boldmath $V_{ub}$ from $B\to \tau\nu$}
 
$B\to \tau \nu$ is an annihilation process. Its BF is related to $V_{ub}$ as:

\begin{eqnarray}
 \label{eq:BR_B_taunu}
{\cal B}(B^{-}\rightarrow\tau^{-}\bar{\nu}_{\tau}) &
= &\frac{G_{F}^{2}m_{B}m_{\tau}^{2}}{8\pi}\left(1-\frac{m_{\tau}^{2}}
{m_{B}^{2}}\right)^{2}\times \nonumber
\\
& & f_{B}^{2}|V_{ub}|^{2}\tau_{B}.
\end{eqnarray}

\begin{figure}[htb]
\centerline{\psfig{file=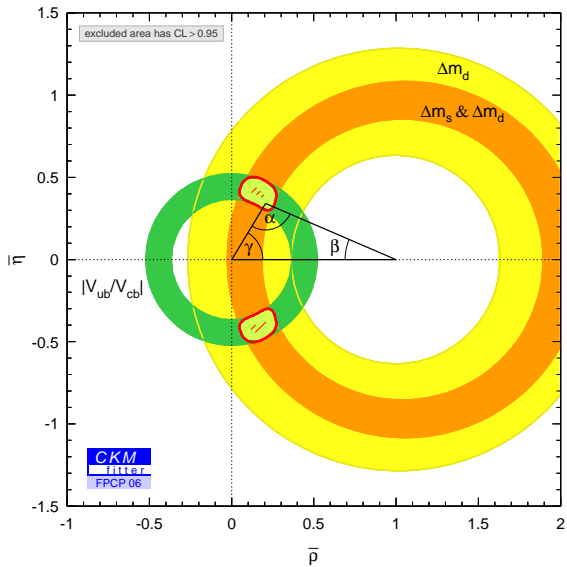,width=7.5cm}}
%\vskip-0.6cm
\caption{CKM fit results for $CP$ conserving observables.}
%\vskip-0.7cm
\label{fig:fig9}
\end{figure}

\noindent
where $G_F$ is the Fermi coupling constant, 
$m_{B}$ and $m_{\tau}$ the $B$ and $\tau$ masses, respectively, 
and $\tau_B$ is the $B^-$ lifetime.
Due to the helicity suppression term,
which is much more severe for muons and electrons, there is practically no
possibility to perform BF measurements of $B\to \mu (e) \nu$ decays at the
B-factories.  
Fully reconstructing one $B$ (hadronic modes for the Belle measurement,
hadronic and semileptonic for the \babar\ measurement) and selecting
candidates for 5 (Belle) or 6 (\babar) $\tau$ decay modes, the energy left
unassigned is mostly from combinatorial background and its distribution
peaks at zero 
for signal. A clear signal with 21 fitted events is observed in Belle's
data~\cite{btaulnubelle}. The measured BF is
$(1.06^{+0.34}_{-0.28}(stat)^{+018}_{-0.16}(syst))10^{-4}$. It has a 4.2 $\sigma$
significance and is in agreement with the SM
expectation~\cite{btaulnubelle2}. Using the HFAG~\cite{hfag} 
average for $V_{ub}$, a first measurement of the $B$ decay constant $f_B$
can be extracted, which is in 
agreement with HPQCD calculations~\cite{hpqcd}.  
\babar\ observes no signal and instead assesses an upper limit for the $B\to
\tau\nu$ BF ($<2.6\time10^{-4}$ at 90\% C.L.)~\cite{btaulnubabar}.  
The ratio of this BF and $\Delta m_d$ constrains $|V_{ub}|/|V_{td}|$.

\begin{figure}[htb]
\centerline{\psfig{file=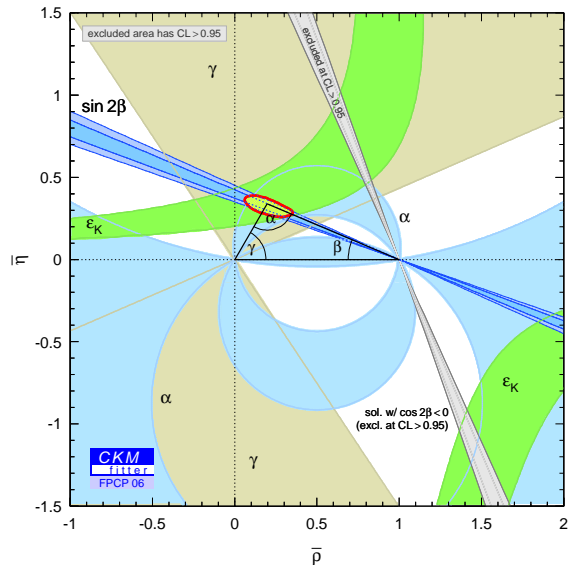,width=7.5cm}}
%\vskip-0.6cm
\caption{CKM fit results for $CP$ violating observables.}
%\vskip-0.7cm
\label{fig:fig10}
\end{figure}

\begin{figure}[htb]
%\vskip-0.6cm
\centerline{\psfig{file=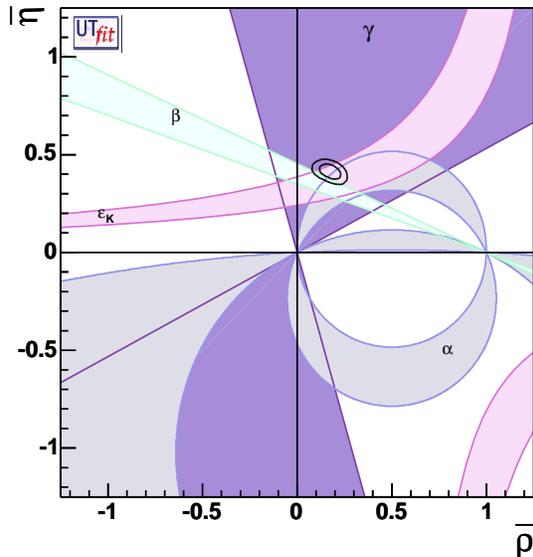,width=8.5cm}}
%\vskip-0.7cm
\caption{1 and 2 $\sigma$ contours of the UT fit results using the side
measurements only compared with the measurements of the single UT angles.}
%\vskip-0.7cm
\label{fig:fig11}
\end{figure}

\section{CONCLUSIONS}

Figs.~\ref{fig:fig9} and~\ref{fig:fig10} show the CKM
fitter results for the $CP$ conserving and $CP$ violating 
observables. The agreement is good in both cases. Fig.~\ref{fig:fig11}
shows the 1 and 2
$\sigma$ contours of the UT fit results with the side
measurements only versus the measurements of the single UT
angles~\cite{utfit}. The 
small disagreement with the $\sin2\beta$ measurement is mostly due to
$V_{ub}$  
inclusive measurements which are almost 3$\sigma$ away from the results of
the fit performed without their inclusion.
All other measurements are, presently, very consistent with
each other and with the SM predictions.

\section{ACKNOWLEDGEMENTS}

I would like to thank the organizers of the BEACH conference for their
help and gracious hospitality in Lancaster. I'd like also to acknowledge
and apologize to all the physicists (theorists and experimentalists) whose
work should be reported here, but could not be for lack of space.


\begin{thebibliography} {99}
\bibitem{charge} Charge conjugate states are also implied.
\end{thebibliography}
\end{document}